\documentclass[amsfonts,amsmath,preprint,superscriptaddress,nofootinbib]{revtex4}

\usepackage{color}
\usepackage{graphicx} 
\usepackage{epstopdf}%

\usepackage{mathpazo}      
\newcommand {\be} {\begin{equation}} 
\newcommand {\ba}{\begin{eqnarray}} 
\newcommand {\ee} {\end{equation}} 
\newcommand{\ea} {\end{eqnarray}}
\newcommand{\eq}[1]{Eq.~(\ref{#1})}

\newcommand {\mn}{{\mu\nu}}

\newcount\hour \newcount\hourminute \newcount\minute 
\hour=\time \divide \hour by 60
\hourminute=\hour \multiply \hourminute by 60
\minute=\time \advance \minute by -\hourminute
\newcommand{\mydate}{\ \today \ - \number\hour :\number\minute}

\preprint{NT@UW-12-11}

\begin{document}

\title{Proton Polarizability Contribution: Muonic Hydrogen Lamb Shift and Elastic Scattering}

\author{Gerald A. Miller}

\affiliation{Department of Physics, Univ. of Washington, Seattle, WA 98195-3560}

\date{\mydate}
\begin{abstract}
The  uncertainty in the contribution to the Lamb shift in muonic hydrogen, $\Delta E^{subt}$ 
 arising  from  proton polarizability effects    in
the two-photon exchange diagram at large virtual photon momenta is shown   large enough to account for the proton radius puzzle. 
This is because   $\Delta E^{subt}$  is determined by an integrand that falls very slowly with  very large virtual photon momenta.    We evaluate the necessary integral using a set of chosen  form factors and also  a dimensional regularization procedure   which makes explicit the need for a low energy constant.  The consequences of our two-photon exchange interaction for low-energy elastic lepton-proton scattering are evaluated and could be observable in a planned low energy lepton-proton scattering experiment planned to   run at PSI. 
 \end{abstract}

\maketitle


\section{Introduction}			\label{sec:one}
\begin{figure}[b]
\bigskip
\begin{center}
\includegraphics[width = 60 mm]{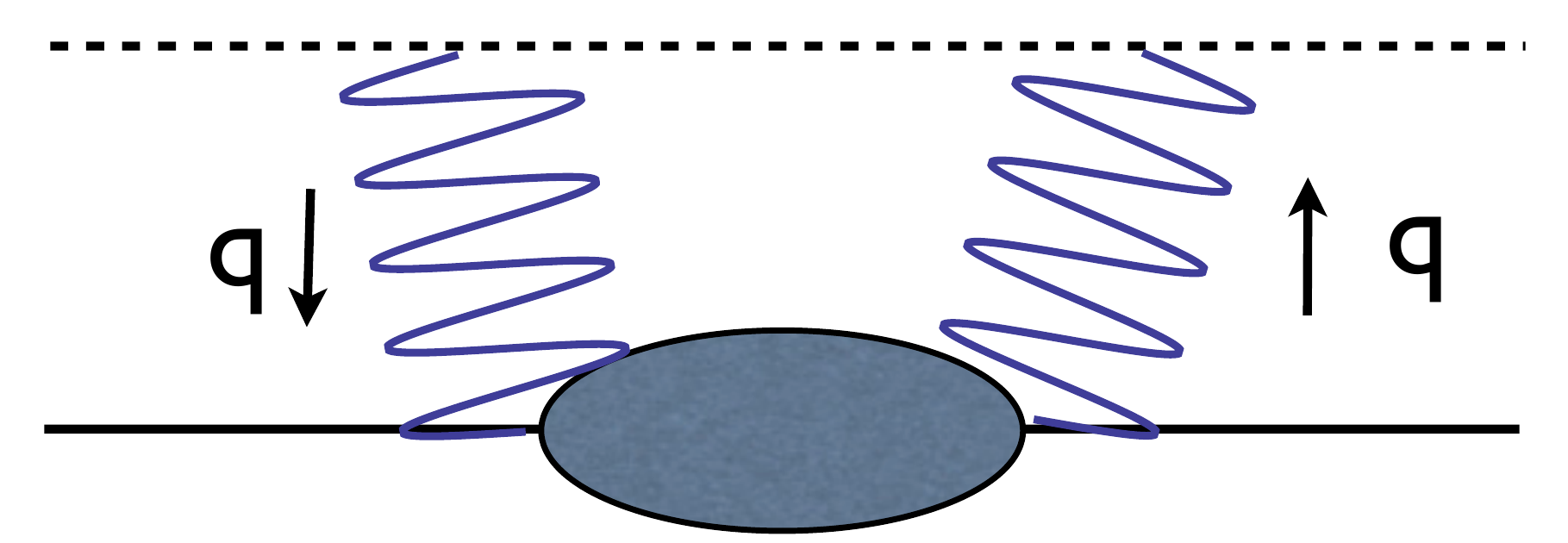}
\caption{The box diagram for the $\mathcal O(\alpha^5m^4)$ corrections. The graph in which the photons cross is also included.
}
\label{fig:lambbox}
\end{center}
\end{figure}
The proton radius puzzle is  one of the most perplexing physics issues of recent times. The 
  extremely precise extraction of the proton radius~\cite{pohl} from the measured 
energy difference between the $2P_{3/2}^{F=2}$ and  $2S_{1/2}^{F=1}$ states of muonic hydrogen disagrees  with that
extracted from electronic hydrogen.
The extracted value of the  proton radius  is smaller than 
the CODATA~\cite{codata} value (based  mainly on electronic H) by about 4\% or 5.0 
standard deviations.  This implies~\cite{pohl} that either the Rydberg constant has to be 
shifted by 4.9 standard deviations or that 
present  QED calculations for hydrogen are insufficient. 
The Rydberg constant is extremely well measured and the 
QED calculations seem to be very extensive and highly accurate, so the muonic H finding is 
a significant puzzle for the entire physics community.

\newcommand{\bea}{\begin{eqnarray}}
\newcommand{\eea}{\end{eqnarray}}
Pohl {\it et al.} show that 
the energy difference
between the  $2P_{3/2}^{F=2}$ and  $2S_{1/2}^{F=1}$ states, $\Delta\widetilde{E}$   is given by
\bea 
\Delta\widetilde{E}=209.9779(49)-5.2262r_p^2+0.0347 r_p^3  \;{\rm meV},\label{rad}
\eea
where $r_p$ is given in units of fm. Using  
this equation and the experimentally measured value $\Delta\widetilde{E}=206.2949$  meV,  one can  see that the difference between the Pohl and CODATA values of the proton radius  
would be   removed by an increase of the first term on the rhs of Eq.~(1) 
by  0.31 meV=$3.1\times 10^{-10}$  MeV.

This proton radius puzzle has been attacked from many different directions~\cite{Jaeckel:2010xx}-\cite{Miller:2011yw}
The present communication is intended to investigate  the  hypothesis that  the proton polarizability contributions entering in the two-photon exchange term,
see Fig.~\ref{fig:lambbox},  can account for the 0.31 meV.
 This idea is worthy of consideration  because the computed effect is proportional to the lepton mass to the fourth power, and 
so is capable of being relevant for muonic atoms, but irrelevant for electronic atoms.

\section{$\Delta E^{subt} $ and its Evaluation    }			\label{sec:two}

The basic idea is that  the two-photon exchange term depends on the forward virtual Compton scattering amplitude $T^{\mu\nu}(\nu, q^2)$  where  $q^2$  is the square of the four momentum, $q^\mu$ of the virtual photon and  $\nu$ is its time component. One  uses  symmetries to  decompose $T^{\mu\nu}(\nu, q^2)$, into  a linear combination of  two  terms,  $T_{1,2}(\nu,q^2)$.  The imaginary parts of   
$T_{1,2}(\nu,q^2)$ are  related to structure functions  $F_{1,2}$ measured in electron- or muon-proton scattering, so that
$T_{1,2}$ can be expressed in terms of $F_{1.2}$ through dispersion relations.  However,
 $F_1(\nu,Q^2)$  falls off too slowly for large values of $\nu$ for the dispersion relation to converge. Hence, one makes  a 
 subtraction at $\nu=0$, requiring that an additional  function of $Q^2$ (the subtraction function) be introduced. One  accounts   for the nucleon  Born terms, and the remainder of  the unknown subtraction function is 
 written as $\overline T_1(0,Q^2)$~\cite{Carlson:2011zd}.  This term is 
   handled   by making a power series expansion around $Q^2=0$, and then using effective field theory 
 to determine the coefficients of the series. The problem with using this expansion  is that  this contribution to the energy is determined by an integral over all values of $Q^2$.
  We proceed by  elaborating the consequences of the behavior of $\overline T_1(0,Q^2)$  for large values of $Q^2$. This is followed by the development of an alternate effective field theory approach to the muon-proton scattering amplitude. In either case, one can account for the needed Lamb shift, while also providing consequences for the two-photon exchange contribution to the scattering amplitude that can be tested in an upcoming experiment~\cite{Arrington:2012}.

\newcommand{\nn}{\nonumber \\&&}

The contribution to the Lamb shift  that is caused by
  $\overline T_1(0,Q^2)$ is denoted as $\Delta E^{subt} $ and is given 
  by~\cite{Pachucki:1996zza,Pachucki:1999zza,Martynenko:2005rc,Carlson:2011zd,Carlson:2011dz}  
\begin{align}&\Delta E^{subt} =  \frac{\alpha^2}{m} \phi^2(0) \int_0^\infty {dQ^2\over Q^2}
	 h({Q^2\over 4m^2})
	 \,\overline T_1(0,Q^2)
					\label{one}\end{align}
where 
$  
 \phi^2(0)={\alpha^3 m_r^3\over 8\pi}$    for the 2S state 
with $m, (m_r)$ as the lepton (reduced) mass, and
\begin{align}
h(t)&= (1-{2t}) \Big((1+{1\over t})^{1/2} - 1\Big) + 1	.\label{three}
\end{align}
The function $h(t) $ is   monotonically falling, approaching $1/\sqrt{t}$ for small values of $t$, and 
falling as $3/(4t)$   large values of $t$.   
 The subtraction function $\overline T_1(0,Q^2)$  is not available from  experimental measurements,  except at the real photon  point $Q^2 = 0$.  It comes from the excitation of the proton, and can   be described, at small values of  $Q^2$, in terms of the electric ($\alpha_E$) and magnetic ($\beta_M$) polarizabilities. For small values of  $Q$ and $\nu=0$ 
 one sees~\cite{Pachucki:1996zza}
$
\lim_{\nu^2,Q^2\to 0}  \overline T_1(0,Q^2) = 
	\frac{Q^2}{\alpha} \beta_M. 	$
 Using this simple linear  $Q^2$-dependence  in \eq{one} shows that  the integral over $\overline T_1(0,Q^2)$ converges at the lower limit, but {\bf diverges logarithmically} at the upper limit. Thus obtaining a non-infinite result depends on   including an arbitrary  form factor that cuts off the integrand for large values of $Q^2$ or some other renormalization procedure.  
 
 We note that $\lim_{Q^2\to\infty}\bar{T}_1(0,Q^2)$ can be obtained from the operator production expansion~\cite{Collins:1978hi,WalkerLoud:2012bg}. Using Eq. (2.18) of  Ref.~\cite{Collins:1978hi}, neglecting the term proportional to light quark masses, and accounting for  different conventions yields $\bar{T}_1(0,Q^2)\sim2.1\; {\rm fm}^{-1}/Q^2$. This $1/Q^2$ behavior removes the putative  logarithmic divergence  of $\bar{T}_1(0,Q^2)$, but this function is far
from determined.

We follow the previous literature by including a
 form factor   defined as    $F_{\rm loop}$. Then 
\be
\overline T_1(0,Q^2) = \frac{\beta_M}{ \alpha} Q^2 F_{\rm loop}(Q^2)	\,.\label{six}
\ee
Using Eqs.~(\ref{one},\ref{three},\ref{six}) one finds the energy shift to be
\begin{align}
&\Delta E^{subt} =  \frac{\alpha^2 \phi^2(0)\; }{ m} {\beta_M\over\alpha} \int_0^\infty  {d Q^2}\left[(1-2 Q^2/(4m^2))\left(\sqrt{1+{4m^2\over Q^2}}-1\right)+1\right]F_{\rm loop}(Q^2).
				\label{de3}		\end{align}
The issue here is the arbitrary nature of the function $F_{\rm loop}(Q^2)$.
Pachucki~\cite{Pachucki:1999zza}  used the dipole form, $\sim 1/Q^4$, often used to characterize 
 the proton electromagnetic form factors. But  the subtraction function should not be computed from the proton form factors, because
 virtual component scattering includes a term in which the photon is absorbed and emitted from the same quark~\cite{Brodsky:1971zh}.
Carlson and Vanderhaeghen~\cite{Carlson:2011zd}  evaluated a loop diagram using a specific model and
found a form factor $\sim 1/Q^2\log Q^2$, leading to a larger contribution to the  subtraction term than previous authors.
Birse \& McGovern~\cite{Birse:2012eb} use terms up to fourth-order in chiral perturbation theory to find
\bea
\label{eq:bm}
\overline T_1^{BM}(0,Q^2) \simeq\frac{\beta_M}{ \alpha} Q^2 \left(1-  {Q^2\over M_\beta^2} +{\cal O}(Q^4)\right) \;\to  
\frac{\beta_M}{ \alpha} Q^2 {1\over\left(1+ {Q^2\over 2M_\beta^2}\right)^2}, \label{bm1}
\eea
with $M_\beta=460 \pm 50 $ MeV.
They also use the most recent evaluation of $\beta_M$, based on  a fit to real 
Compton scattering~\cite{Griesshammer:2012we} that   finds
\be
\beta_M =  (3.1 \pm 0.5) \times 10^{-4} {\rm\ fm}^3,\label{betam}
\ee
where only statistical and Baldin Sum Rule errors are included. Their  result is a negligible 
 $\Delta E^{subt}= 4.1\mu $ eV~\cite{Birse:2012eb}.  
The form \eq{bm1} achieves the correct $1/Q^2$ asymptotic behavior of $ \overline T_1(0,Q^2) $ but the coefficient 
$\beta_M/\alpha$ is not the same as obtained from the operator product expansion. The coefficient of \eq{bm1} is about twice the asymptotic limit obtained by Collins~\cite{Collins:1978hi}.

Previous  authors~\cite{Carlson:2011zd,Birse:2012eb} noted the sensitivity of the integrand of \eq{de3} to large values of $Q^2$. 
Our aim here is to more  fully explore the uncertainty in the subtraction term that arises from the logarithmic divergence.
We shall  use a form of $F_{\rm loop}(Q^2)$  that is consistent with the  constraint on the $Q^4$ term found Birse \& McGovern~\cite{Birse:2012eb}. This is done  by postulating a term that begins at order $Q^6$ in \eq{six}, such as
 \be
F_{\rm loop}(Q^2)=\left({Q^2\over M_0^2}\right)^n{ 1\over (1+ a Q^2)^N },\; n\ge2,\;N\ge n+3,\label{mine}\ee
where $M_0,a$  are  parameters to be determined.  With \eq{mine} 
the low $Q^2$ behavior of $\bar{T}_1(0,Q^2)$ is of order $Q^6$ or greater and it falls as $1/Q^4$ or greater for large values of $Q^2$. 
So far as we know, there are no constraints on the coefficient of the $Q^6$  term and the $1/Q^4$ term. However, we shall determine the subtraction term's contribution to the  Lamb shift as a general function of $n,N$.  We note that $\beta_M$ is anomalously small due to a cancellation between pion cloud and intermediate $\Delta$ terms~\cite{Thomas:2001kw} , so that one can use a value ten times larger than appears in \eq{betam} to set the overall scale of the subtraction term.
Thus we replace the term $\beta_M$ of \eq{six} by a general form of the same  dimensions $\beta$:
$ \beta_M\rightarrow \beta.$

 The use of \eq{mine} in \eq{de3} allows one to state the expression for the energy shift in closed form as a general function of
 $n,N.$ We find
\bea&&  \Delta E^{subt} =  \frac{\alpha^2 \phi^2(0)\; }{ m} {\beta\over\alpha} \left({1\over a M_0^2}\right)^n J_{n,N}(m^2a), \\&&
J_{n,N}(m^2a)\equiv {1\over a} \int_0^\infty\;dx\;{x^n\over (1+x)^N}\left[\left(1-{x\over 2m^2 a}\right)\left((1+{4m^2a\over x})^{1/2}-1\right)+1\right] .
\label{jeq}\eea
The integral over $x$ can be obtained in a closed form in terms of hypergeometric functions. However, a much more understandable expression can be obtained by replacing the bracketed expression in \eq{jeq} by its large argument limit $(3m^2a/x)$. This approximation is valid
over the entire range of the integrand because of the presence of the factor $x^n$ with $n\ge 2$.
Then one obtains
\bea  J_{n,N}(m^2a)\approx 3m^2
 {\Gamma(N-n)\Gamma(n)\over \Gamma(N)}=3m^2 B(N,n),
\label{jeq2}\eea
so that 
\bea \Delta E^{subt} \approx 3 {\alpha^2 m\phi^2(0)\; }  {\beta \over\alpha} \lambda^n  B(N,n),\;\lambda\equiv{1\over M_0^2a}.\label{myde}\eea
Numerical evaluations show that the approximation is accurate to  better than a quarter of a percent.
The expression \eq{myde} makes clear the $m^4$ dependence of the contribution to the Lamb shift.

 The numerical value of the  term $ \Delta E^{subt} $ depends on $(n,N),\beta$ and the combination $M_0^2a\equiv \lambda^{-1}$:
\bea \Delta E=3.91 {\rm meV \;fm^3} \beta \lambda^n B(N,n).\eea 
If we take $N=5,n=2$ so that $B(5,2)=1/12$, and $\beta =10^{-3}$ fm$^{-3}$, a  value of $\lambda= 30.9$ reproduces 
$ E=0.31\; {\rm meV}$. If we take $M_0=0.5 $ GeV (as in \cite{Birse:2012eb}) ,  then $a^{-1}=15.4 \;{\rm GeV}^2$, and 
 that the contribution to the integral comes from the region of very high values of $Q^2$. Other values of
$n,N$ and $\lambda$ could be used to get the identical contribution to the Lamb shift.

 Chiral perturbation theory could be used to determine the terms of order $Q^6$ and higher in $ \overline T_1^{BM}(0,Q^2) $,
but this procedure is always limited to a finite number of terms. Indeed one could use values of $n$ greater than 2, and still reproduce the needed contribution to the Lamb shift.

The above discussion shows that the current procedure used to estimate the size of the subtraction term is rather arbitrary. This arises because the chiral EFT is being applied to the virtual-photon nucleon scattering amplitude. Another technique    would be to develop an effective field theory  to determine the short-distance lepton-nucleon amplitude implied by the subtraction term.  

\section{Effective field theory for the $\mu p$ interaction}
The previous considerations show that the value of  $ \Delta E^{subt} $  depends heavily
on assumptions  the behavior of $\overline T_1(0,Q^2),\; (F_{|rm loop}$) for large values 
$Q^2$. This is true even though the leading $1/Q^2$ term is known.  The underlying cause of this uncertainty  is the would-be logarithmic divergence in the integral of \eq{de3} for the case $F_{\rm loop}=1$. This is a symptom that some other  technique could be  used~\cite{Georgi:1994qn}.
 Another  way to proceed is to use an effective field theory (EFT) for the
lepton-proton interaction~\cite{Caswell:1985ui}. 
In EFT, logarithmic divergences identified through dimensional regularization  are  renormalized away by including   a lepton-proton contact interaction in the Lagrangian.

We may handle the divergence using
 standard dimensional regularization 
  (DR) techniques by  evaluating the scattering amplitude of Fig.~1.
  The term of interest is obtained by including only  $\overline T_1(0,Q^2)$ of \eq{three} with $F_{\rm loop}=1$.
 We evaluate the loop integral in  $d=4-\epsilon$ dimensions and obtain the result:
\bea {\cal M}_2^{DR}({\rm loop}) ={3\over2}i\; \alpha^2 m{ \beta_M\over \alpha}\big[{2\over \epsilon}+\log {{\mu^2\over m^2}}+{5\over 6}-\gamma_E+\log 4\pi\big]\overline u_f  u_i  \overline  U_f U_i ,\label{res1}\eea
where lower case spinors represent  leptons of mass $m$, and upper case proton of  mass $M$,  $q$ is momentum transferred to the proton, and $\gamma_E$ is Euler's constant, 0.577216$\cdots\;$.

The result \eq{res1} corresponds to an infinite contribution to the Lamb shift in the limit that $\epsilon$ goes to zero. In EFT one removes the divergent piece by adding a lepton-proton contact interaction to the Lagrangian that removes the divergence, replacing it by an unknown finite part. 
The finite part is obtained by    fitting  to a relevant  piece of data.
Here the only relevant data is the 0.31 meV needed to account for the proton radius puzzle. 
The low energy term  contributes
\bea {\cal M}_2^{DR}({\rm LET})=i C(\mu), \label{LET}\eea where  $C(\mu)$ is chosen  such that the sum of the terms of 
\eq{res1} and \eq{LET},  $ \equiv   {\cal M}_2^{DR}$, is finite and independent of the value of $\mu$.
Thus we  write the resulting scattering amplitude as 
\bea {\cal M}_2^{DR} =i\; \alpha^2 m{ \beta_M\over \alpha}(\lambda +5/4)\;\overline u_f  u_i  \overline  U_f U_i ,\label{mdr}
 \eea 
where $\lambda $ is  determined by fitting to the Lamb shift. \eq{mdr}  corresponds to using the $\overline{MS}$ scheme because  
the term $\log (4\pi)-\gamma_E$  is absorbed into $\lambda$. 

The corresponding contribution to the Lamb shift is 
given by
\bea \Delta E^{DR}=\alpha^2 m{ \beta_M\over \alpha}\phi^2(0)(\lambda +5/4).\label{de4}\eea
Setting $ \Delta E^{DR}$ to 0.31 meV in the above equation requires that $\lambda =769$, which  seems like a large number. However, as noted above,   $\beta_M$ is extraordinarily small~\cite{Thomas:2001kw}.  The natural units of polarizability are ${\beta_M\over \alpha}\sim
 4\pi/\Lambda_\chi^3$,~\cite{Butler:1992ci}
  where $\Lambda_\chi \equiv 4\pi f_\pi$, ($f_\pi$ is  the pion decay constant). 
Then \eq{mdr}  becomes
 \bea {\cal M}_2^{DR} =i\; 3.95 \;\alpha^2 m {4\pi\over \Lambda_\chi^3 }\overline u_f  u_i  \overline  U_f U_i .\label{mdr1}\eea
The coefficient 3.95 is  of natural  size.
Thus standard EFT techniques result in an effective lepton-proton interaction of natural size that is proportional to the lepton mass. 
The form of \eq{mdr1} is not unique. There are other possible operators that reduce to that form in the low-energy, low-momentum regime
of relevance here.

The present results, \eq{myde} and \eq{de4} represent an assumption that there is a  lepton-proton interaction of standard-model origin, caused by the 
 high-momentum behavior of the virtual scattering amplitude, that is sufficiently large to account for the proton radius puzzle. 
 Fortunately, our hypothesis can be   tested in  an upcoming  low-energy  $\mu^\pm p, e^\pm p$ scattering experiment~\cite{Arrington:2012}   planned to occur at  PSI.  


\section{  Lepton Proton Scattering at Low Energies }

Our aim is to determine the consequences of the particular two-photon exchange term for lepton-proton scattering at low energies.
Our previous attempt~\cite{Miller:2011yw} implied very large corrections to quasi-elastic electron-nucleus scattering that are severely in disagreement with experiment~\cite{Miller:2012ht}. It is necessary to check that a similar large unwanted contribution does not
appear here. Thus we 
  provide a prediction for the PSI experiment.  It is well-known that two-photon exchange effects in electron-proton scattering  are small
at low energies. Our contact interaction is proportional to the lepton mass, so  it could provide a measurable  effect for muon-proton scattering but be ignorable for electron-proton scattering. 
We shall investigate the two  consequences of using form factors (FF) and effective field theory (DR). 
\newcommand{\kslash}{\not\hspace{-0.7mm}k}\newcommand{\lslash}{\not\hspace{-0.7mm}l}
\newcommand{\pslash}{\not\hspace{-0.7mm}p}
\newcommand{\sslash}{\not\hspace{-0.7mm}s}

The invariant amplitude is given as  $ {\cal M}_{fi}\equiv {\cal M}_{fi}^{(1)}+ {\cal M}_{fi}^{(2)}$
where  the superscripts denote the number of photons exchanged. 
The first  term is given by 
\bea {\cal M}_{fi}^{(1)} =\mp i {e^2\over q^2+i0}\overline u_f\gamma_\mu u_i  \overline U_f \Gamma^\mu U_i,
\eea
where $u_{i,f}$ represent  leptons of mass $m$, $U_{i,f}$  represent the proton of  mass $M$, and $q$ is momentum transferred to the proton.
  The minus sign holds for negatively charged muons, and the plus sign for positively charged muons.
\bea \Gamma^\mu=\gamma^\mu F_1(q^2)+ i{\sigma^\mn \over 2M}q_\nu F_2(q^2),\eea
with $P_f=P_i+q$.
The second-order term that arising  from the use  of \eq{six} and \eq{mine} , which uses form factors  (FF)  is given by 
\bea {\cal M}_{fi}^{(2)} (FF)=-{e^4\over (2\pi)^4}\int d^4 k{1\over k^2+i0} {\overline T_1(0,-k^2)\over (k-q)^2+i0}\overline u_f L u_i \overline U_f     U_i, \label{m1}
\eea
where
\bea&&
L 
= {2mk^2+4p_i\cdot k\;\kslash \over -4(p_i\cdot k)^2 +k^4+i0}.
\label{l1}
\eea
The second-order amplitude arising from the use of EFT is given above in \eq{mdr}.
 
The cross section depends upon the average over initial and sum over final fermion spins, as denoted by an over line. We first obtain $ \overline{ \left | {\cal M}_{fi}^{(1)}\right|^2}$. Standard text-book expressions use the Mott form, obtained by
ignoring the lepton mass. This is not a good approximation for the muons of the experiment~\cite{Arrington:2012} which 
have momenta ranging between 100 and 200 MeV/c, and our  terms of \eq{m1} and \eq{mdr}  would vanish in that approximation.
We find
\bea&&
 \overline{ \left| {\cal M}_{fi}^{(1)}\right|^2}
=[16M^2 (\varepsilon\varepsilon'+q^2/4) (F_1^2-{q^2\over 4M^2}F_2^2 ) +4G_M^2(q^4/2+q^2m^2)]
\left({e^2\over q^2+i0 }\right)^2, \nn
\label{m1sq}\eea
where $\theta$ is the laboratory scattering angle, and $\varepsilon (\varepsilon')$ is the  incident (final)  lepton laboratory total energy.
  \begin{figure}[t]
\begin{center}
\includegraphics[width = 160 mm]{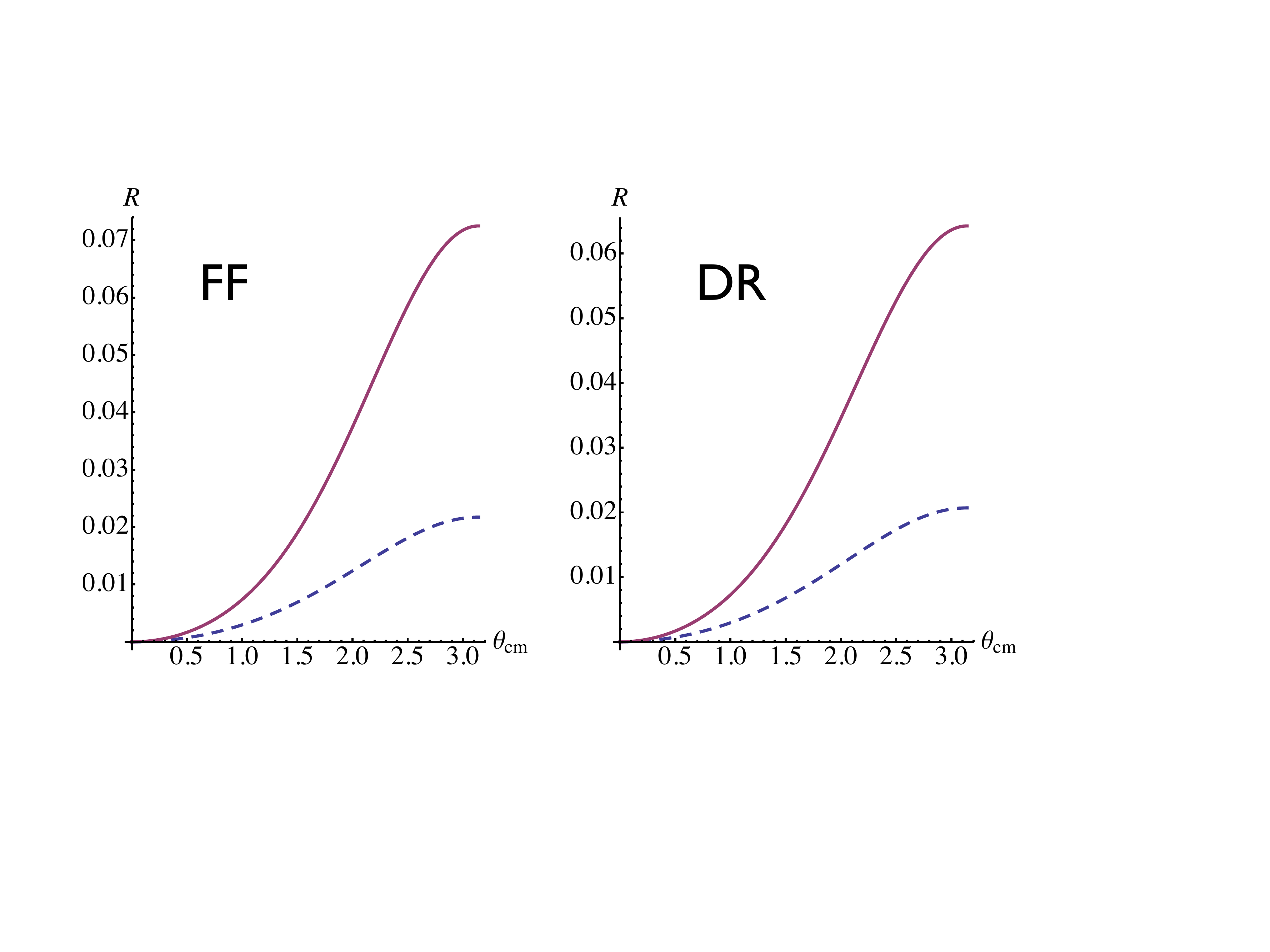}
\caption{The ratio $R$ obtained using form factor (FF) regularization or dimensional regularization (DR).
The solid curves show the results for muon laboratory  momentum of 200 MeV/c and the dashed curves show the results for 100 MeV/c.}
\label{fig:R}
\end{center}
\end{figure}
The present  interference term of interest $\Delta\equiv  2{\rm Real}\;\overline{ [( {\cal M}_{fi}^{(1)})^*{(\cal M}_{fi}^{(2)})]}$
is obtained using standard trace algebra. We find
\bea&&\Delta(FF)=
 {8M G_E(q^2)}{ \mp i e^2\over q^2+i0}{e^4\over (2\pi)^4}\int d^4 k{1\over k^2+i0} {\overline T_1(0,-k^2)\over (k-q)^2+i0} {1\over(-4(p_i\cdot k)^2 +k^4+i0)} \nn\times[2m^2k^2(4\varepsilon M +q^2) +4p_i\cdot k\;k\cdot(P_i+P_f)q^2/2
+2p_i\cdot kk\cdot(p_i+p_f)(4\varepsilon M+q^2 )].
\eea
We have seen that the integrand is dominated by large values of $k$, therefore we neglect $q$ in the integrand. This allows considerable simplification so that we find
\bea&&\Delta(FF)={\mp96G_E(q^2)}M^2{  e^2\over q^2+i0} \varepsilon\alpha {\beta}m^2 \lambda^n B(N,n) \label{dff}\eea
 where the terms $\lambda,B(N,n)$ appear in \eq{myde}.
A negligible  term  proportional to the square of the incident lepton momentum has been dropped. 
The term  $\Delta$ adds to the square of the lowest order term for $\mu^- p$ interactions, as expected from an attractive interaction that increases the Lamb shift.   The computed value of $\Delta(FF)$ does not depend on $n,N,\lambda$ for those values that
reproduce the needed Lamb shift via \eq{myde}.

For EFT the contribution to the cross section via interference can be worked out using \eq{mdr1} to be
\bea&&\Delta^{DR}={\mp 8 [4\varepsilon M+q^2] \alpha ( \lambda+{5\over4}) m^2 \beta_M G_E(q^2)}M{  e^2\over q^2+i0}.\label{ddr}\eea

We are now prepared to display the effects of our two-photon exchange term on $\mu^--p$ scattering at low energies.
The size of the effect is represented by the ratio $R$, with
\bea 
R\equiv {\Delta\over  \overline{ \left| {\cal M}_{fi}^{(1)}\right|^2}}.\label{rat}\eea
The  ratio $R>0$ for $\mu-p $ scattering. The numerator of \eq{rat} is obtained from either \eq{dff} (FF) or \eq{ddr} (DR).
 The ratio $R$ is proportional to the square of the lepton mass, which is negligible for $e^\pm-p$ scattering.
 We consider two muon  momenta 100 and 200 MeV/c. The results are shown in Fig.~\ref{fig:R}. The angular dependence is dominated by the $Q^2=-q^2$ term inherent in  \eq{rat}. The two sets of curves are very similar because the size of the effect
 is constrained by the required energy shift of 0.31 meV. The size of the effect should be detectable within the expected  sub-1 \% accuracy of the PSI experiment. We emphasize that our calculation is valid only at  low muon laboratory energies. 
 \section{Summary and Discussion} 				\label{sec:end}

The findings of this paper can be summarized with a few statements:
\begin{itemize}

\item The integrand (see \eq{one}) that determines the value of $\Delta E^{subt} $ values slowly with large values of $Q^2$,
causing  the uncertainty in the evaluation to be   large enough to account for the proton radius puzzle.

\item The integrand can be evaluated using one of an infinite set of possible form factors or a dimensional regularization procedure.

\item Either method can be used to account for the proton radius puzzle and predict an observable effect of a few percent
for low energy $\mu-p$ scattering.

\end{itemize}

The literature \cite{pohl}-\cite{Miller:2011yw} poses several explanations for the proton radius puzzle: The electronic-hydrogen experiments might not be as accurate as previously reported, $\mu-e$ universality might be violated, and that  a strong interaction effect entering in a loop diagram  is important for muonic hydrogen, but not for electronic hydrogen. It is beyond the scope of the present paper to argue for the unique  correctness of any one of these ideas. The strong-interaction effect discussed here is large enough to be  testable experimentally,


\begin{acknowledgments}
I thank M. J. Savage, B. Long, R. Pohl, S. J. Brodsky, G. Paz, R. Hill, M. Birse, and R. Gilman  and A. W. Thomas for useful discussions. This research was supported by the United States Department of Energy,   grant FG02-97ER41014.   I   gratefully acknowledge the support and gracious hospitality of the University of Adelaide  during the formative stages of this work.
\end{acknowledgments}



\end{document}